\documentstyle[12pt,psfig]{article}
\begin{document}
\begin{center}
\medskip
\medskip
\medskip
\medskip
\medskip
\medskip
\medskip
\Large{\bf Determination of equation of state of quark matter from 
 $J/\psi$ and $\Upsilon$ suppression at RHIC and  LHC}
\vskip 0.2in

\large{Dipali Pal, Binoy Krishna Patra, Dinesh Kumar Srivastava}
\vskip 0.2in

\normalsize{ Variable Energy Cyclotron Centre\\
 1/AF Bidhan Nagar, Calcutta 700 064\\
 India\\}

\vskip 0.2in

Abstract

\vskip 0.2in

\end{center}

The long life-time of the quark-gluon plasma likely to be created
in the relativistic heavy ion collisions at RHIC and LHC energies 
renders it sensitive to the details of the equation of state
of the quark-matter. We show that the $p_T$ dependence of
the survival probability of the directly produced $J/\psi$ at
   RHIC energies and that
of the  directly produced $\Upsilon$ at LHC energies is quite sensitive to the
speed of sound in the quark matter, which  relates the pressure
and the energy density of the plasma. The transverse expansion of the
plasma is shown to strongly affect the $J/\psi$ suppression at
LHC energies.

\vskip 0.3in


\newpage
\section{Introduction}
Statistical quantum chromodynamics predicts that at sufficiently high densities
or temperatures the quarks and gluons confined inside hadrons undergo a
deconfining phase transition to a plasma of quarks and gluons.
The last two decades of high energy nuclear physics  activity has been 
directed towards the production of
this new state of matter through relativistic heavy ion collisions.
This has led to experiments at BNL AGS
 and CERN SPS and to the building of the BNL Relativistic Heavy Ion Collider 
and a planning of the ALICE experiment at the CERN Large Hadron Collider.
With the reported confirmations of the quark-hadron phase transition at
the  relativistic heavy ion collision experiments at the CERN SPS~\cite{qm99}, 
the first step in the 
search for quark-gluon plasma, which pervaded the early
universe, microseconds after the big bang and which may be present in the
core of neutron stars, is complete.

 The emphasis of the experiments at the BNL RHIC and the CERN LHC will now
  necessarily shift to an accurate determination of the properties
of the quark matter.  An important observable for this is the speed of
sound in the plasma, defined through:
\begin{equation}
c_s^2=\frac{\partial p}{\partial \epsilon}.
\end{equation}
Often one writes,
\begin{equation}
p=c_s^2 \epsilon
\label{eos}
\end{equation}
for the equation of state of the quark matter, where
$\epsilon$ is the energy density and $p$ is the pressure. For the simplest
bag-model equation of state (with $\mu_B=0$), we write
\begin{eqnarray}
\epsilon&=&3 a T^4 + B,\\
p&=&a T^4-B,\\
a&=&\left[2 \times 8 +\frac{7}{8}\times 2 \times 2 \times 3 \times N_f
\right ] \frac{\pi^2}{90}~,
\end{eqnarray}
with $B$ as the bag pressure and $N_f$ as the number of flavours, so that
$c_s^2=1/3$. In general $\Delta=\epsilon-3p$ measures the deviation of 
the equation
of state from the ideal gas of massless quarks and gluons
(when it is identically zero)
and depends sensitively on the interactions present in the plasma. The
lattice QCD calculations show that $\Delta\geq$~0, till the temperature
is several times the critical temperature~\cite{laer}. This implies
that in general $c_S^2 \leq 1/3$. Any experimental information on this
will be most welcome. 

We show in the present work that the transverse momentum dependence of
the survival probability of the $J/\psi$ and $\Upsilon$ at RHIC and
LHC energies are quite sensitive to the value of the speed of sound. The
very long life-time of the plasma likely to be attained at LHC makes it even
more sensitive to the details of the equation of state  of the quark
matter through the transverse expansion of the plasma.

\section{Formulation}
The theory of quarkonium suppression in QGP~\cite{ms} is very well studied and
several excellent reviews exist~\cite{ramona1,helmut1,helmut2},
 which dwell both on the phenomenology as well
as on the experimental situation. We recall the basic details
which are relevant for the present demonstration. 

The interquark potential for (non-relativistic) quarkonium states
at zero temperature may be written as:
\begin{equation}
V(r,0)=\sigma\, r -\frac{\alpha}{r}
\label{v0}
\end{equation}
where $r$ is the separation between $Q$ and $\overline{Q}$. The
bound-states of $c\overline{c}$ and $b\overline{b}$ are well described
if the  parameters
$\sigma=$ 0.192 GeV$^2$, $\alpha=$0.471, $m_c=$ 1.32 GeV, and $m_b=$~4.746~~GeV
are used~\cite{mehr}. At finite temperatures the potential is modified
due to colour screening, and evolves to:
\begin{equation}
V(r,T)=\frac{\sigma}{\mu (T) }\left[ 1 - e^{-\mu(T)r} \right] 
-\frac{\alpha}{r}e^{-\mu(T) r}.
\end{equation}
The screening mass increases with temperature. When $\mu(T) \rightarrow 0$,
the equation (\ref{v0}) is recovered. At finite temperature,
 when $r\rightarrow 0$ the $1/r$ behaviour is dominant, while as 
$r\rightarrow \infty$ the range of the potential decreases with $\mu(T)$.
This makes the binding less effective at finite temperature. Semiclassically,
one can write for the energy of the pair,
\begin{equation}
E(r,T)=2m_Q+\frac{c}{m_Q r^2}+V(r,T)
\end{equation}
where $<p^2><r^2>=c=\cal{O}(1)$. Radius of the bound state at any temperature
is obtained by minimizing $E(r,T)$. Beyond some critical value $\mu_D$
for the screening mass $\mu(T)$, no minimum is found. The screening is now
strong enough to make the binding impossible and the resonance can not
form in the plasma. The ground state properties of some of the
quarkonia reported by authors of Ref.~\cite{mehr} are given in table~1.
We have also listed the formation time of these resonances defined in
Ref.\cite{ks} as the time taken by the heavy quark to traverse a
distance equal to the radius of the quarkonium in its rest frame 
$\sim m_Q r_{Q\overline{Q}}/p_{Q\overline{Q}}$, where  $p_{Q\overline{Q}}$
is the momentum of either of the quarks of the resonance. It may be recalled
that somewhat different values for the formation time are reported by
Blaizot and Ollitrault~\cite{blaz} who solve the bound-state problem within
the WKB approximation and define the formation time as the time spent
by a quark in going between the two classical turning points.

Now let us consider a central collision in a nucleus-nucleus collision, which
results in the formation of quark gluon plasma at some time $\tau_0$.
 Let us concentrate  at $z=0$ and on the region of energy density,
 $\epsilon \geq \epsilon_s$ which encloses the
plasma which is dense enough to cause the melting of a particular state
of quarkonium.  We assume the plasma to cool, according
to Bjorken's boost invariant (longitudinal) hydrodynamics and then
generalize our results to include the transverse expansion of the
plasma. We assume that the $Q\overline{Q}$ pair is produced at the
transverse position $\mathbf{r}$ at
$\tau=0$ on the $z=0$ plane with momentum $\mathbf{p}_T$. In the collision 
frame, the pair would take a time equal to $\tau_F E_T/M$ for the
quarkonium to form, where $E_T=\sqrt{p_T^2+M^2}$ and $M$ is the mass of
the quarkonium. During this time, the pair would have moved to
 the location $(\mathbf{r}+\tau_F \mathbf{p}_T/M)$. If at this instant, the
plasma has cooled to an  energy density less than $\epsilon_s$, the pair
would escape and quarkonium would be formed. If however, the energy density
is still larger than $\epsilon_s$, the resonance will not form and
we shall have a quarkonium suppression~\cite{blaz,chu,kp}.

It is easy to see that the $p_T$ dependence of the survival probability
will depend on how rapidly the plasma cools. If the initial energy density
is sufficiently high,  the plasma will take longer to cool and only the
pairs with very high $p_T$ will escape. If however the plasma cools rapidly,
then even pairs with moderate $p_T$ will escape. The transverse expansion
of the plasma can further accelerate the rate of cooling giving us an 
additional handle to explore the equation of state, which as we know,
will control the expansion of the plasma.

\subsection{Longitudinal expansion of the plasma}

As indicated,  we  first take the Bjorken's boost-invariant longitudinal
 hydrodynamics
to explore the expansion of the plasma. Thus, the energy momentum tensor
of the plasma is written as~\cite{bj};
\begin{equation}
T^{\mu\nu}=(\epsilon+p)u^\mu u^\nu + g^{\mu \nu} p,
\label{tmunu}
\end{equation}
where $\epsilon$ is the energy-density, $p$ is the pressure, and $u^\mu$ is the
four velocity of the fluid, in a standard notation. If the effects of
viscosity are neglected, the energy-momentum conservation is
given by
\begin{equation}
\partial_\mu T^{\mu \nu}=0.
\label{econs}
\end{equation}
The assumption of the boost-invariance provides that the energy density,
pressure, and temperature become functions of only the proper time $\tau$
and that the Eq.(\ref{econs}) simplifies to
\begin{equation}
\frac{d\epsilon}{d\tau}=-\frac{\epsilon+p}{\tau}.
\label{bj}
\end{equation}

The effect of the speed of sound is seen immediately. Using the 
Eq.(\ref{eos}), we can now write
\begin{equation}
\epsilon(\tau) \tau^{1+c_s^2}= \epsilon(\tau_0)\tau_0^{1+c_s^2}=\mbox{const.}
\end{equation}
so that if $c_s^2$ is small, the cooling is slower. Chu and Matsui~\cite{chu}
explored the consequence of the extremes $c_s^2=1/3$ and $c_s^2=0$ 
on the $p_T$ dependence of the survival probability. We shall explore
the sensitivity of the quarkonium suppression to the equation of
state by (some-what arbitrarily) choosing two values of the speed
of sound, $1/\sqrt{3}$ and $1/\sqrt{5}$, in the following. 

We now have all the ingredients to write down the survival probability
and we closely follow Chu and Matsui for this.

We take a simple parametrization for the energy-density profile:
\begin{equation}
\epsilon(\tau_0,r)=\epsilon_0\left[1-\frac{r^2}{R^2}\right]^{\beta} \theta(R-r)
\end{equation}
where $r$ is the transverse co-ordinate and $R$ is the radius of the nucleus.
One can define an average energy density $<\epsilon_0>$ as
\begin{equation}
\pi R^2 <\epsilon_0>=\int 2\pi\, r \, dr \epsilon(r)
\end{equation}
so that 
\begin{equation}
\epsilon_0=(1+\beta) <\epsilon_0>.
\end{equation} 
We have taken $\beta=1/2$, which may be thought as indicative of
the energy deposited being proportional to the number of participants
in the system. In case one feels that the energy deposited may be proportional
to number of nucleon-nucleon collisions then one can repeat the calculations
with $\beta=1$, which will reflect the proportionality of the deposited
energy  to the nuclear thickness. The average energy-density is obtained
from the Bjorken formula:
\begin{equation}
<\epsilon_0>=\frac{1}{\pi R_T^2 \tau_0}\,\frac{dE_T}{dy}
\end{equation}
where $E_T$ is the transverse energy deposited in the collision.

The time $\tau_s$ when the energy density drops to $\epsilon_s$ is easily 
estimated as
\begin{eqnarray}
\tau_s(r)&=&\tau_0 \left[ \frac{\epsilon(\tau_0,r)}{\epsilon_s}\right]
         ^{1/(1+c_s^2)}\nonumber\\
         &=&\tau_0\left[\frac{\epsilon_0}{\epsilon_s}\right]^{1/(1+c_s^2)}
            \left[ 1-\frac{r^2}{R^2}\right]^{\beta/(1+c_s^2)}
\end{eqnarray}

As discussed earlier~\cite{chu}, we can equate the duration of screening
$\tau_s(r)$ to the formation time $t_F=\gamma \tau_F$ for the quarkonium
to get the critical radius, $r_s$:
\begin{equation}
r_s=R\left[ 1-\left( \frac{\gamma \tau_F}{\tau_{s0}}\right)^{(1+c_s^2)/\beta}
      \right]^{1/2}\, \theta\left [ 1-\frac{\gamma \tau_F}{\tau_{s0}}\right],
\end{equation}
where $\tau_{s0}=\tau_s(r=0)$.  This critical radius, is seen to mark
the boundary of the region where the quarkonium formation is suppressed.
As discussed earlier, the quark-pair will escape the screening region (and
 form quarkonium) if its position and transverse
momentum $\mathbf{p}_T$ are such that
\begin{equation}
\left| \mathbf{r}+\tau_F \mathbf{p}_T/M\right| \geq r_s.
\end{equation}
Thus, if $\phi$ is the angle between the vectors $\mathbf{r}$ and 
$\mathbf{p}_T$,
 then 
\begin{equation}
\cos \phi\,\geq\,\left[(r_s^2-r^2)\,M-\tau_F^2\,p_T^2/M\right]/
                 \left[2\,r\,\tau_F\,p_T\right],
\label{phi}
\end{equation}
which leads to a range of values of $\phi$ when the quarkonium would
escape. We also realize that if the right hand side of the above equation
is greater than 1, then no angle is possible when the quarkonium can
escape. Now we can write for the survival probability of the quarkonium:
\begin{equation}
S(p_T)=\left[\int_0^R \, r \, dr \int_{-\phi_{\mbox{max}}}
                                     ^{+\phi_{\mbox{max}}}\,
          d\phi\, P(\mathbf{r},\mathbf{p}_T)\right]/
        \left[2\pi \int_0^R \, r\, dr\, P(\mathbf{r},\mathbf{p}_T)\right],
\label{spt}
\end{equation}
where $\phi_{\mbox{max}}$ is the maximum positive angle ($0\leq \phi \leq \pi$)
allowed by Eq.(\ref{phi}), and
\begin{equation}
\phi_{\mbox{max}}=\left\{ \begin{array}{ll}
                   \pi & \mbox{if $y\leq -1$}\\
                   \cos^{-1} |y| & \mbox{if $-1 < y < 1$}\\
                   0          & \mbox{if $y \geq 1$}
                          \end{array}
                   \right .,
\end{equation}
where
\begin{equation}
y= \left[(r_s^2-r^2)\,M-\tau_F^2\,p_T^2/M\right]/
                 \left[2\,r\,\tau_F\,p_T\right],
\end{equation}
and $P$ is the probability for the quark-pair production at $\mathbf{r}$
with transverse momentum $\mathbf{p}_T$, in a hard collision. Assuming that
the $\mathbf{p}_T$ and $\mathbf{r}$ dependence for hard collisions factor out,
we approximate
\begin{equation}
P(\mathbf{r},\mathbf{p}_T)=P(r,p_T)=f(r)g(p_T),
\end{equation}
where we take
\begin{equation}
f(r)\propto \left[ 1-\frac{r^2}{R^2}\right]^\alpha \theta(R-r)
\end{equation}
with $\alpha=1/2$. The Eq.(\ref{spt}) can be solved analytically for some
limiting cases of $p_T$ etc., see Ref.~\cite{chu}.

\section{Transverse expansion of the plasma}

It is generally accepted that the rare-faction wave from the surface
of the plasma will reach the centre by $\tau=R/c_s$. For the case of lead
nuclei, this comes to about 12 fm/$c$. If the life time of the QGP 
is comparable 
to this time, the transverse expansion of the plasma can not be ignored.
The transverse expansion of the plasma will lead to a much more rapid
cooling than suggested by a purely longitudinal expansion.

For the four-velocity of the collective flow we write:
\begin{equation}
u^\mu=(\gamma, \gamma \mathbf{v})
\end{equation}
where $\mathbf{v}$ is the collective flow velocity and $\gamma=1/\sqrt{1-v^2}$.
We further assume that the longitudinal flow of the plasma has a scaling
solution, so that the boost-invariance along the longitudinal direction
remains valid. Assuming cylindrical symmetry, valid for central collisions,
it can be shown that the four-velocity $u^\mu$ should have the form,
\begin{equation}
u^\mu=\gamma_r(\tau,r)(t/\tau,v_r \cos \phi, v_r \sin \phi, z/\tau),
\end{equation}
with
\begin{eqnarray}
\gamma_r&=&\left[ 1 - v_r^2(\tau,r)\right]^{-1/2}\nonumber\\
\tau&=&(t^2-z^2)^{1/2}\nonumber\\
\eta&=&\frac{1}{2}\ln \frac{t+z}{t-z}.
\end{eqnarray}

Thus all the Lorentz scalars are now functions of $\tau$ and $r$, and 
independent of the space-time rapidity $\eta$. This reduces the $(3+1)$
dimensional expansion with cylindrical symmetry and boost-invariance
along the longitudinal direction to:
\begin{equation}
\partial_\tau T^{00}+r^{-1}\partial_r(rT^{01})+\tau^{-1}(T^{00}+p)=0
\end{equation}
and
\begin{equation}
\partial_\tau T^{01}+r^{-1}\partial_r\left[r(T^{00}+p)v_r^2\right]+
   \tau^{-1}T^{01}
+\partial_r p=0
\end{equation}
where
\begin{equation}
T^{00}=(\epsilon+p)u^0u^0-p
\end{equation}
and 
\begin{equation}
T^{01}=(\epsilon+p)u^0 u^1.
\end{equation}

We see that the speed of sound which will appear through the dependence
of the pressure $p$ on the energy-density will affect the 
time-evolution of all the quantities, especially through the gradient 
terms. This is of-course extensively  documented. 
We solve these equations using well established methods~\cite{hydro}
with initial energy density profiles as before and
estimate the constant energy density contours\cite{jane,munshi} 
appropriate for $\epsilon=\epsilon_s$
to get $\tau_s(r)$. Rest of the treatment follows as before. In these
calculations we have assumed the initial transverse velocity to be
identically zero.

We only need to identify the initial conditions. We consider $Pb+Pb$
collisions ($Au+Au$, for RHIC) with the initial average energy densities:
\begin{equation}
<\epsilon_0>=\left\{ \begin{array}{ll}
                   6.3 \mbox{~~GeV/fm$^3$} & \mbox{SPS, $\tau_0=0.5$~fm}\\
                    &\\
                   60 \mbox{~~GeV/fm$^3$} & \mbox{RHIC, $\tau_0=0.25$~fm}\\
                     &\\
                   425 \mbox{~~GeV/fm$^3$} & \mbox{LHC, $\tau_0=0.25$~fm}\\
                          \end{array}
                   \right .
\end{equation}

The estimate for SPS is obtained from assumption of QGP formation in $Pb+Pb$
 experiments, while those for RHIC and LHC are taken from the
self-screened parton cascade calculation~\cite{sspc}.
If a formation time for the SPS energies is assumed to be of the
order of 1 fm/$c$, the estimate given above will drop to about 3 GeV/fm$^3$.
A larger initial energy density than the one assumed here for RHIC and
LHC could be obtained by using the concept of parton saturation~\cite{kari}.
We are, however, interested in only a demonstration of the effect
of equation of state on the quarkonium suppression, and we feel that
the values used here are enough for this.

\section{Results}
\subsection{Speed of sound vs. transverse expansion}

It is quite clear that a competition between the speed of sound and the
onset of the transverse expansion during the life-time of the deconfining
matter can lead to interesting possibilities. In order to illustrate this
and to explore the consequences, we show in Fig.~1 the  time
corresponding to the constant energy
density contours which enclose the deconfining matter- which can 
dissociate the directly produced $J/\psi$ (see table~1), at RHIC energies. 
We see that  owing to the (relatively) short time that the QGP would take
to cool down to $\epsilon_s^{J/\psi}$, the effect of the transverse flow
is marginal and for $c_s^2=1/3$, limited to large radii. Large changes 
in the contour are seen when the speed of sound is varied. This is very
important indeed. Note that the cooling to the value appropriate for 
$\Upsilon$ suppression is attained too quickly to be affected by the
transverse expansion, and even the change due to variation in the speed
of sound is quite small. Of-course the duration of the deconfining 
medium is prolonged if the speed of sound is reduced.

The corresponding results for the LHC energies are shown in Fig.~2.
Now we see that at $r=0$ the duration of the deconfining medium
reduces by a factor of $2$ when the speed of sound is $1/\sqrt{3}$, and
the transverse expansion of the plasma is allowed. This is a consequence
of the longer time which the plasma takes to cool at LHC.

The scenario for the $\Upsilon$ dissociating matter at LHC 
is quite akin to the case of $J/\psi$ dissociating medium at RHIC;
the results are affected by the speed of sound and not by the
transverse flow (Fig.~3).

\subsection{Consequences for survival of quarkonia}

Now we return to the transverse momentum dependence of the
survival probabilities. As a first step, we plot the
survival of the directly produced $J/\psi$'s at SPS, RHIC and LHC
energies (Fig.~4) when only longitudinal expansion is accounted for
and the speed of sound is varied. We see that RHIC energies provide
the most suitable environment to measure the speed of sound with the
help of $J/\psi$ suppression. The variations in the $p_T$ dependence
is too meagre at SPS (due to  a very short duration of the deconfining
medium) and at LHC (now, due to  a very long duration!) when the 
speed of sound is varied.

This advantage of RHIC energies is maintained when the transverse 
expansion is accounted for (Fig.~5). As one could have expected from the
contours (Fig.~1), the results are more sensitive to the variation
of the speed of sound than to the transverse flow. The accuracy of
the procedure is seen from the fact that the survival probability
around $p_T=$ 15 GeV for $c_s^2=1/3$ is identical for the longitudinal
and the transverse expansion of the plasma. This is a direct reflection
of the identity of the corresponding contours near $r=0$ (Fig.~1).

The $J/\psi$ suppression at LHC energies, as indicated, becomes 
sensitive to the transverse flow, the shape of the survival probability
changes and the largest $p_T$ for which the formation is definitely
possible is enhanced by about 10 GeV (Fig.~6). 

The $\Upsilon$ suppression, which in our prescription is possible
only at the LHC energy, is seen to be clearly affected by the
speed of sound but not by the transverse expansion of the plasma (Fig.~7).

\subsection{Consequences of chain decays of quarkonia}

The entire discussion so far has been in terms of the directly produced
$J/\psi$'s and $\Upsilon$'s. However it is well established that only 
about 58\% of the observed $J/\psi$ in $pp$ collisions originate
directly, while 30\% of them come from $\chi_c$ decay and 12\% from
the decay of $\psi^\prime$. Thus the survival probability of the $J/\psi$
in the QGP can be written as:
\begin{equation}
S=0.58\,S_{\psi}+0.3\,S_{\chi_c}+0.12\,S_{\psi^\prime}~,
\end{equation}
in an obvious notation. We give the survival probabilities of these
resonances for a transversely expanding plasma at RHIC with the
speed of sound as $1/\sqrt{3}$ in the left panel of Fig.~8. We see
that the  competition of the formation times and the duration of
the sufficient dissociation
energies render a rich detail to the suppression pattern of the
charmonia.  We shall later argue
that the $\psi^\prime$'s are easily destroyed by a moderately hot
hadronic gas as well, as their binding energy is on the order of
just 50 MeV. The right panel of the figure shows 
the survival probability as a function of the transverse momentum
when the speed of sound in the plasma is varied from $1/\sqrt{3}$ to
$1/\sqrt{5}$. We see that while the gross features for shape of the
survival probability remain similar to Fig.~5,  as seen earlier,
the survival of the $J/\psi$ for larger $p_T$ is now enhanced as
the $\chi_c$'s, which decay to form $J/\psi$ start escaping. 
Thus the complete escape of the $\chi_c$ having $p_T\,>$ 10 GeV 
(for $c_s^2=1/3$) lends a distint kink in $S(p_T)$ at $p_T \approx$ 10 GeV.
Its location, which shifts to about 14 GeV when the speed of sound 
is decreased, can perhaps be more accurately determined
by plotting the derivative $dS/dp_T$,
which will have a discontinuity there. If the statistics is
really good (which unfortunately is a somewhat difficult proposition),
this discontinuity in the $dS/dp_T$ for the $J/\psi$ can be a
unique signature of melting of the resonance in plasma,
as the $p_T$ dependence of the survival probability 
due to absorption by hadrons should be weak~\cite{hadron} and
 smooth~\cite{helmut1}.

The corresponding results for the LHC energies are given in Fig.~9, in
an analogus manner, and we see the characteristic `kink' in $S(p_T)$
has shifted to about $p_T=16$ GeV, announcing the complete escape of
$\chi_c$ from the plasma.

The decay contribution of the resonances completely alters the shape of
the survival probabilities for $\Upsilon$ from that seen earlier (Fig.~7).

We note that (Fig.~10) both $\chi_b$ and $\Upsilon^\prime$ get suppressed
at the RHIC energies, while the directly produced $\Upsilon$ is likely
to escape. However, as only about 54\% of the $\Upsilon$'s may be directly
produced, while about 32\% have their origin in the decay of $\chi_b$ and
a further 14\% in the decay of $\Upsilon^\prime$, the resultant (right
panel, Fig.10) survival probability is just about 50\% for the
lowest $p_T$, signalling the suppressions of the higher resonances.

The results for LHC energies become quite dramatic (Fig.~11)
as now the `kink' in the survival probability is very clearly
seen at $p_T \sim $20 GeV for $c_s^2=1/3$ and at $\sim$ 26 GeV for the lower speed
of sound. If one looks at the $dS/dp_T$ 
then it could be very useful indeed. We may add that fluctuations 
in the initial conditions etc. may perhaps make it difficult to
notice this aspect for $J/\psi$'s, though for the $\Upsilon$'s
these should survive, provided we have good statistics at these $p_T$.

The large difference in this behaviour seen between charmonium and bottomonium
 suppression has its origin in the large difference in the energy
 densities required to melt the different resonances of charmonia and bottomonia. 

\subsection{Absorption by nucleons and comovers}

So far we have discussed the fate of quarkonia only  when the
presence of quark gluon plasma is considered. It is very well
established that there are several aspects like initial state
scattering of the partons, shadowing of partons, absorption of
the pre-resonances ( $|Q\overline{Q}g>$ states) by the nucleons
before they evolve into physical quarkonia, and also dissociation
of the resonances by the comoving hadrons ~\cite{helmut1,helmut2}.
It has been argued that the absorption by co-moving hadrons
will be important for $\psi^\prime$, due to its very small binding
energy, while for more tightly bound resonances it may be 
weak~\cite{helmut1}.

Let us briefly comment on them one-by-one. Shadowing of partons
should play an important role in the reduced production of quarkonia,
especially at the LHC energies. It is clear that if shadowing is
important, we shall witness a larger effect on $J/\psi$ than on $\Upsilon$,
because of the smaller values of the $x$ for gluons. At the same time, the
effect of shadowing should be similar for different resonances of the
charmonium (or bottomonium), as similar $x$ values would be involved for 
them. 

The absorption of the pre-resonances by the nucleons is another
source of $p_T$ dependence. It is important, to recall once
again that as the absorption is operating on the pre-resonance, the
effect should be identical for all the states of the quarkonium
which are formed.  

This is a very important consideration as it is clear that if we 
look  at the ratio of rates for different states of $J/\psi$ or
the $\Upsilon$ family as a function of $p_T$,
 then in the absence of QGP-effects they would
be identical to what one would have expected in absence of nuclear
absorption and shadowing, providing a clear pedestal for the observation
of QGP~\cite{ramona2}.

There is another aspect of $p_T$ dependence which needs to be 
commented upon, before we conclude. The (initial state) scattering of
partons, before the gluons of the projectile and the target nucleons fuse
to produce the $Q\overline{Q}$-pair, leads to  an increase of the
$<p_T^2>$ of the resonance which emerges from the collision\cite{sour}.
The increase in the $<p_T^2>$, compared to that for $pp$ collisions
is directly related to number of collisions the nucleons are
likely to undergo, before the gluonic fusion takes place. This leads to
a rich possibility of relating the average transverse momentum of the
quarkonium to the transverse energy deposited in the collision
(which decides the number of participants and hence the number of collisions).
Considering that collisions with large $E_T$ may have formation of QGP
in the dense part of the overlapping region, the quarkonia, which are
produced in the densest part (and hence contributing the largest
increase in the transverse momentum) are also most likely to melt
and disappear. This may lead to a characteristic saturation and even
turn-over of the $<p_T^2>$ when plotted against $E_T$ when the
QGP formation takes place. In absence of QGP, this curve would
continue to rise with $E_T$.

Obviously, all these (well explored and yet non-QGP) effects need to
be accounted for, before we can begin to see the suppression of the
quarkonium due to the formation of QGP. It seems that this has 
been achieved at least at the SPS energies~\cite{helmut1,helmut2}.

The next step would obviously be the one discussed in the present work,
that of looking for the $p_T$ dependence of the survival probability,
to see if we can get more detailed information on the equation of the
state of the plasma. This would require high precision data, extending
to larger $p_T$ at several $E_T$. This may prove to be difficult,
though not impossible in principle at least. It will however prove
to be very valuable, if it can be done.

\section{Summary and Discussion}

We have seen that the survival probability of $J/\psi$ at RHIC
energies and that of $\Upsilon$ at LHC energies can provide valuable 
information about the equation of state of the quark matter, as the
results are not affected by uncertainties of transverse expansion
of the plasma. If the transverse expansion of the plasma takes
place, it gives a distinct shape to the survival probability of the
$J/\psi$ at LHC energies, whose detection will be a sure signature of
the transverse flow of the plasma within the QGP phase.

Before concluding we would add that in these exploratory demonstrations
we have chosen some specific values~\cite{ks} for the deconfining matter
which can dissociate quarkonia. These could be different, in particular
the $\epsilon_s^{J/\psi}$ could be much larger than the value
used here.  This, however, will not change the basic results as the time-scales
involved in the $\Upsilon$ and the $J/\psi$ suppression are so
very different.

The other uncertainty comes from the usage of $\epsilon_s$ as the criterion
for deconfinement. One could have as well used the Debye mass as the
temperature and the fugacity changed in a chemically equilibrating
plasma, to fix the deconfining zone. This can indeed be done,
along with the other extreme of the Debye mass estimated from
lattice QCD. This has been studied in great detail by authors
of Ref.~\cite{ramona2} and can be easily extended to the present case.
We plan to do it in a future publication.

In brief, we have shown that the $J/\psi$ and $\Upsilon$ suppression
at RHIC and LHC energies can be successfully used to map the
equation of state for the quark-matter. As the two processes will map
different but over-lapping regions, taken together, these results
will help us to explore a vast region of the equation of state.

{\em We thank  Bikash Sinha for useful discussions.}

\newpage
\medskip
\begin{table}[h]
\caption{ Critical screening masses, etc. 
           for quarkonia~\protect\cite{mehr,ks}.}
\vskip 0.2in
\centerline{
\begin{tabular}{|l|l|l|l|l|l|l|} 
\hline
  & & & & & &   \\
 State & $J/\psi$ & $\chi_c$ & $\psi^\prime$ &$\Upsilon$ 
 &$\Upsilon^\prime$ & $\chi_b$ \\
  && & & & &   \\
\hline
& && & & & \\ 
$M$  (GeV) & 3.1 & 3.5& 3.7 & 9.4 & 10.0 & 9.9 \\
& && & & &  \\
\hline
& && & & & \\ 
$r$ (fm) & 0.45 & 0.70 &0.88& 0.23 & 0.51 & 0.41 \\
& && & & &  \\
\hline
& && & & & \\ 
$\tau_F$ (fm) & 0.89 & 2.0 &1.5& 0.76 & 1.9 & 2.6 \\
& && & & &  \\
\hline
& && & & & \\ 
$\mu_D$ (GeV) & 0.70 & 0.34 &0.36& 1.57 & 0.67 & 0.56 \\
& && & & &  \\
\hline
& && & & & \\ 
$T_d/T_c$ & 1.17 & 1.0 &1.0& 2.62 & 1.12 & 1.0 \\
& && & & &  \\
\hline
& && & & & \\ 
$\epsilon_s$ (GeV/$fm^3$) & 1.92 & 1.12&1.12 & 43.37 & 1.65 & 1.12 \\
& && & & &  \\
\hline
\end{tabular}
}
\end{table}
\newpage
\begin{figure}
\psfig{file=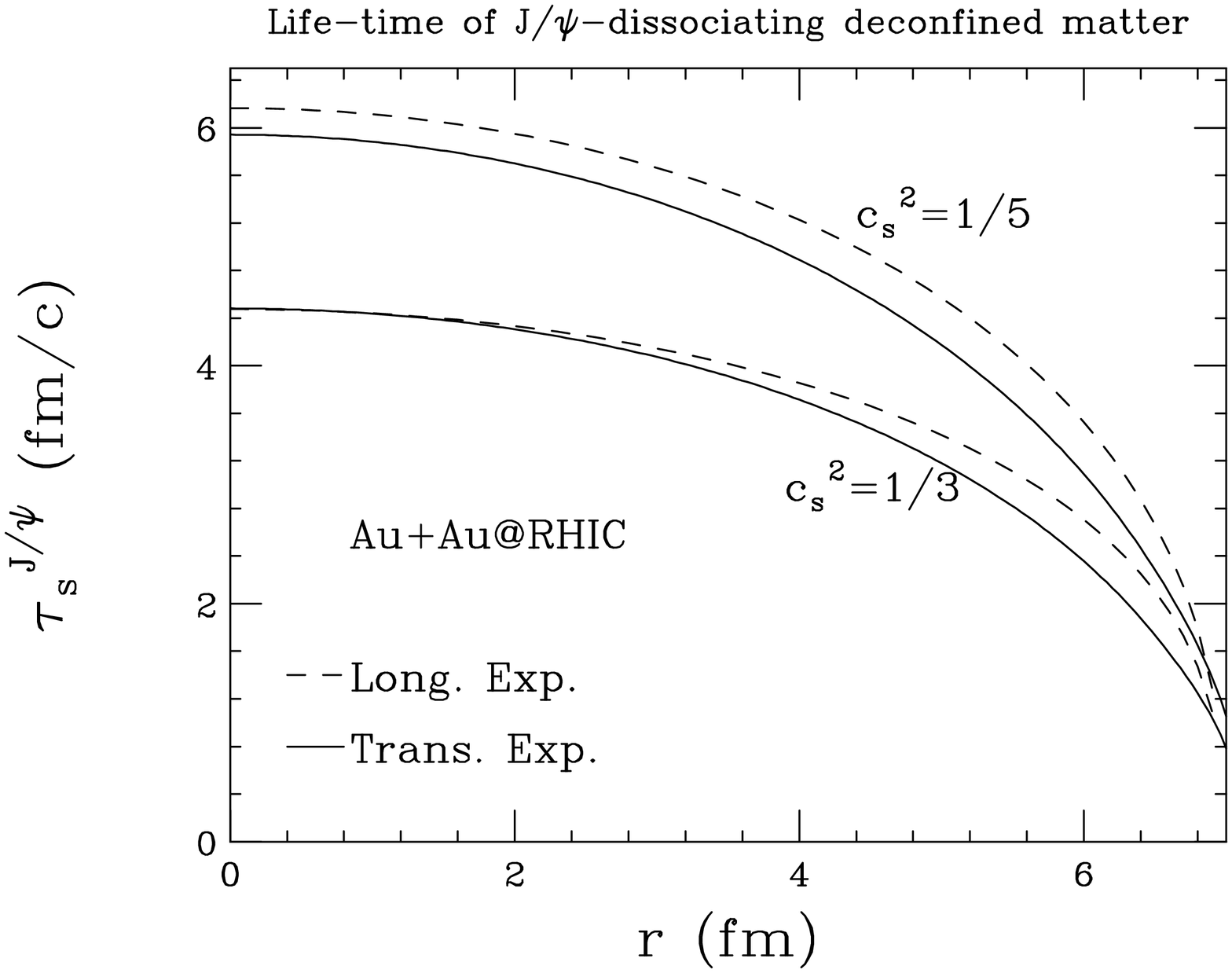,angle=90,height=12cm,width=15cm}
\vskip 0.1in
\caption{The contour of deconfining matter capable of suppressing
direct $J/\psi$ formation at RHIC energies. The solid curves
are results with transverse flow, while the dashed curves are for 
longitudinal flow.
}
\end{figure}
\newpage
\begin{figure}
\psfig{file=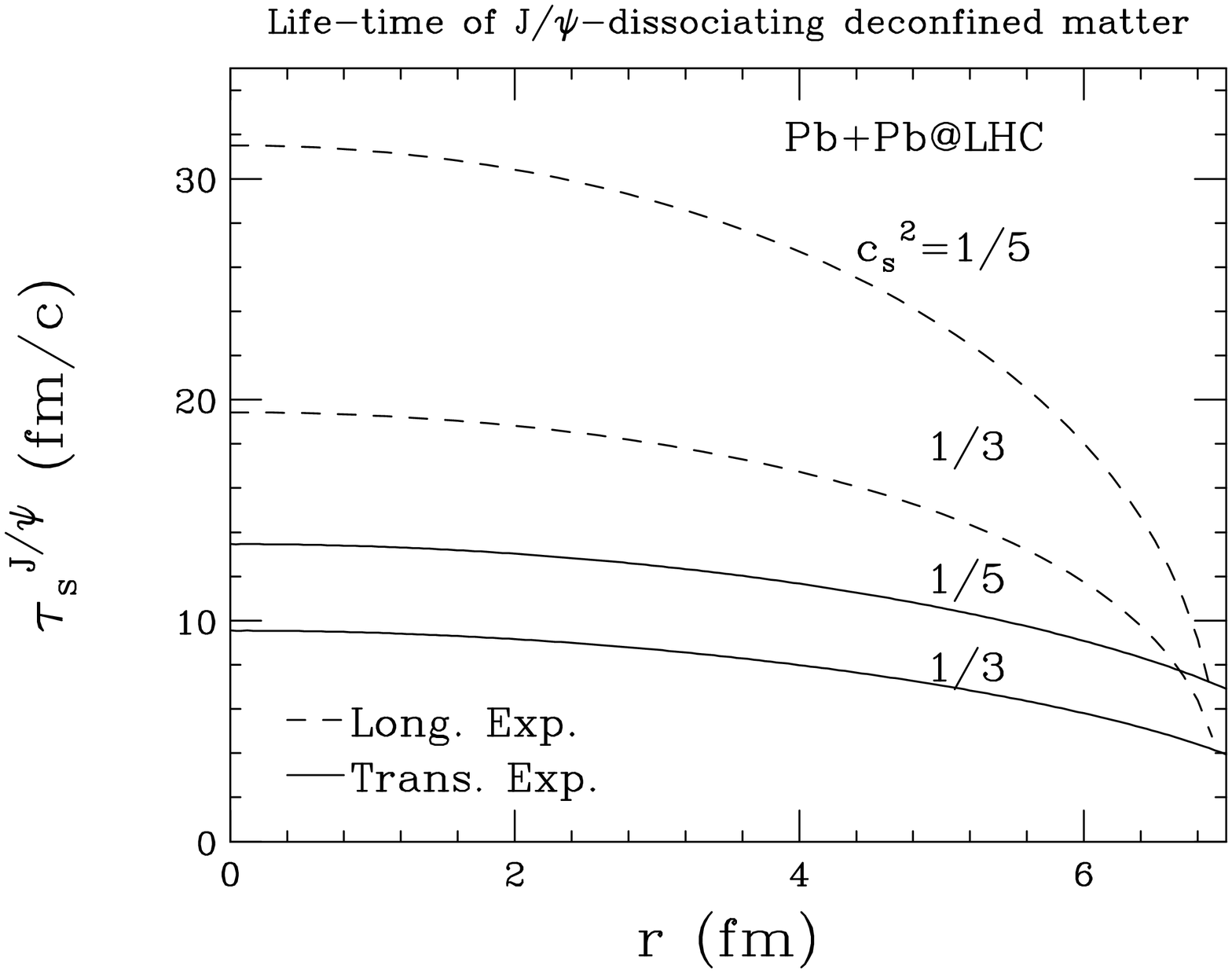,angle=90,height=12cm,width=15cm}
\vskip 0.1in
\caption{The contour of deconfining matter capable of suppressing
direct $J/\psi$ formation at LHC energies. The solid curves
are results with transverse flow, while the dashed curves are for 
longitudinal flow.
}
\end{figure}
\newpage
\begin{figure}
\psfig{file=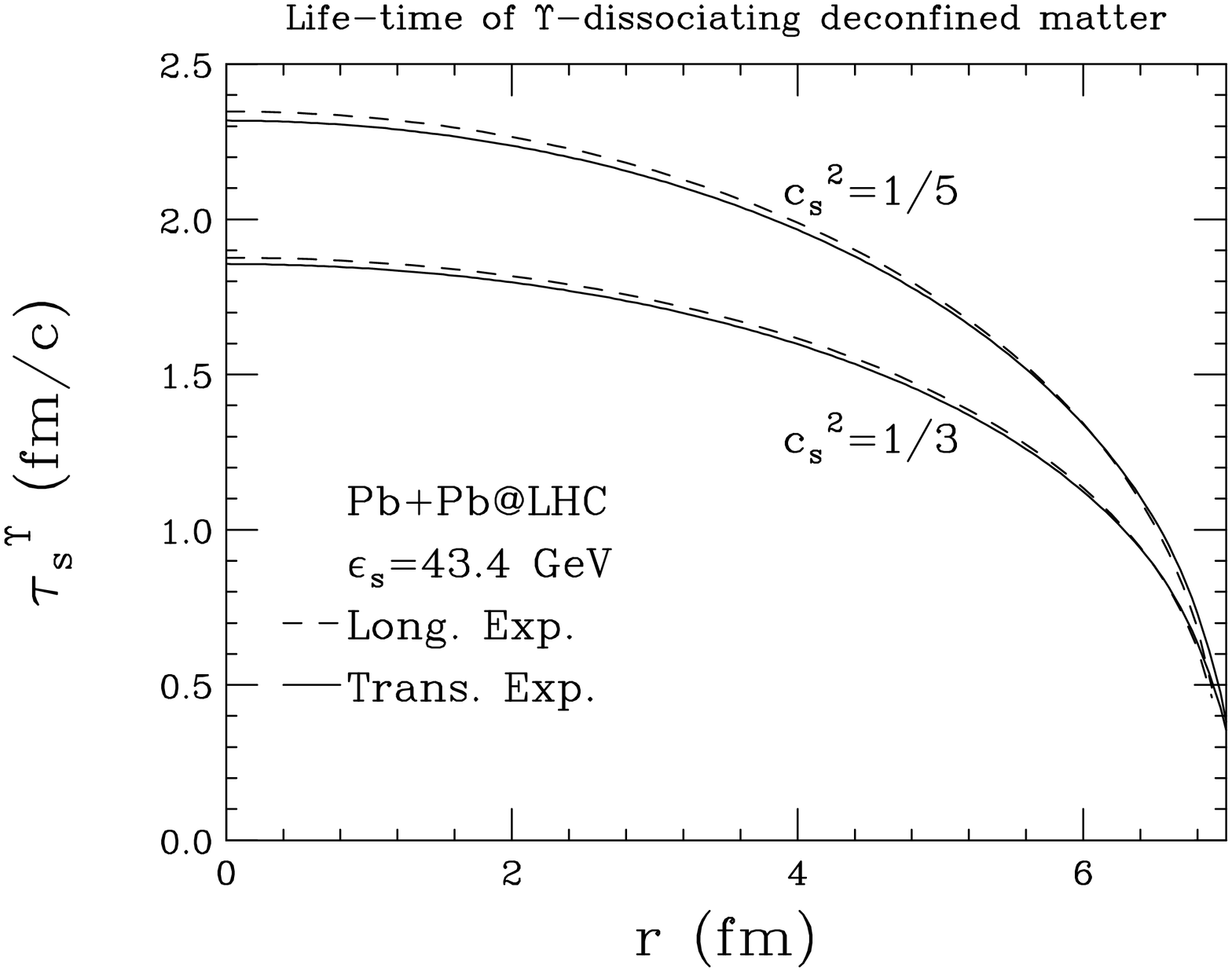,angle=90,height=12cm,width=15cm}
\vskip 0.1in
\caption{The contour of deconfining matter capable of suppressing
direct $\Upsilon$ formation at LHC energies. The solid curves
are results with transverse flow, while the dashed curves are for 
longitudinal flow.
}
\end{figure}
\newpage
\begin{figure}
\psfig{file=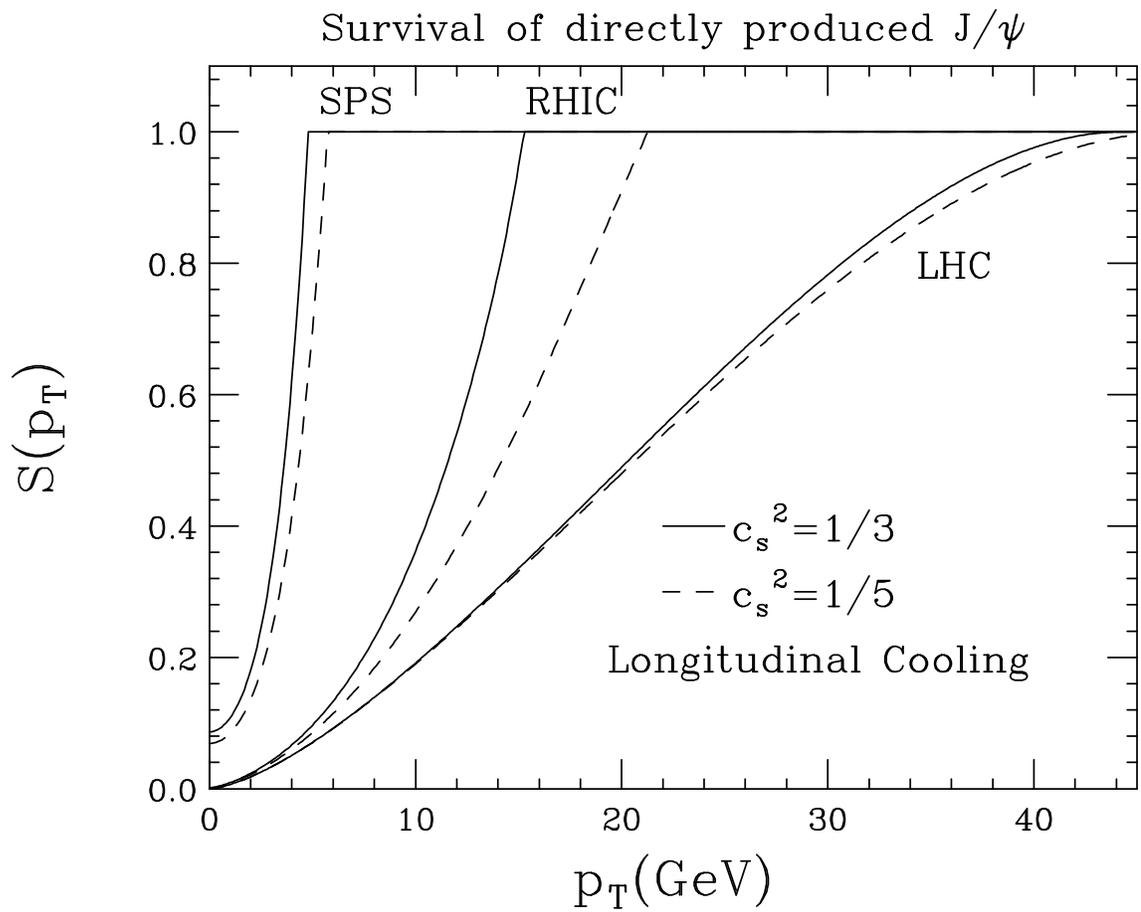,angle=90,height=12cm,width=15cm}
\vskip 0.1in
\caption{The survival probability for $J/\psi$ at SPS, RHIC, and
LHC energies, with only longitudinal cooling, when speed of sound
is changed.
}
\end{figure}
\newpage
\begin{figure}
\psfig{file=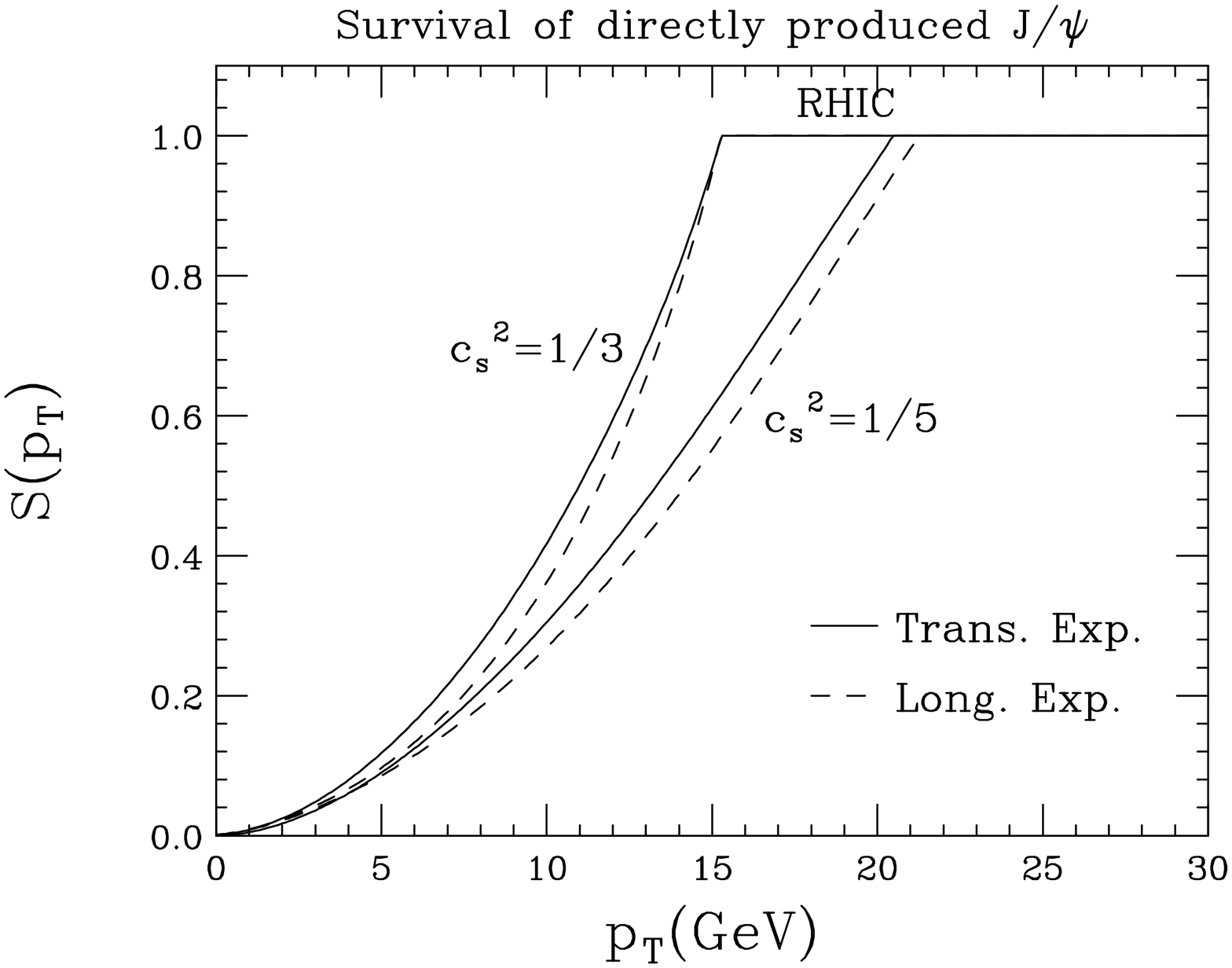,angle=90,height=12cm,width=15cm}
\vskip 0.1in
\caption{The survival probability for $J/\psi$ at  RHIC energy
with longitudinal and transverse cooling when speed
of sound is changed.
}
\end{figure}
\newpage
\begin{figure}
\psfig{file=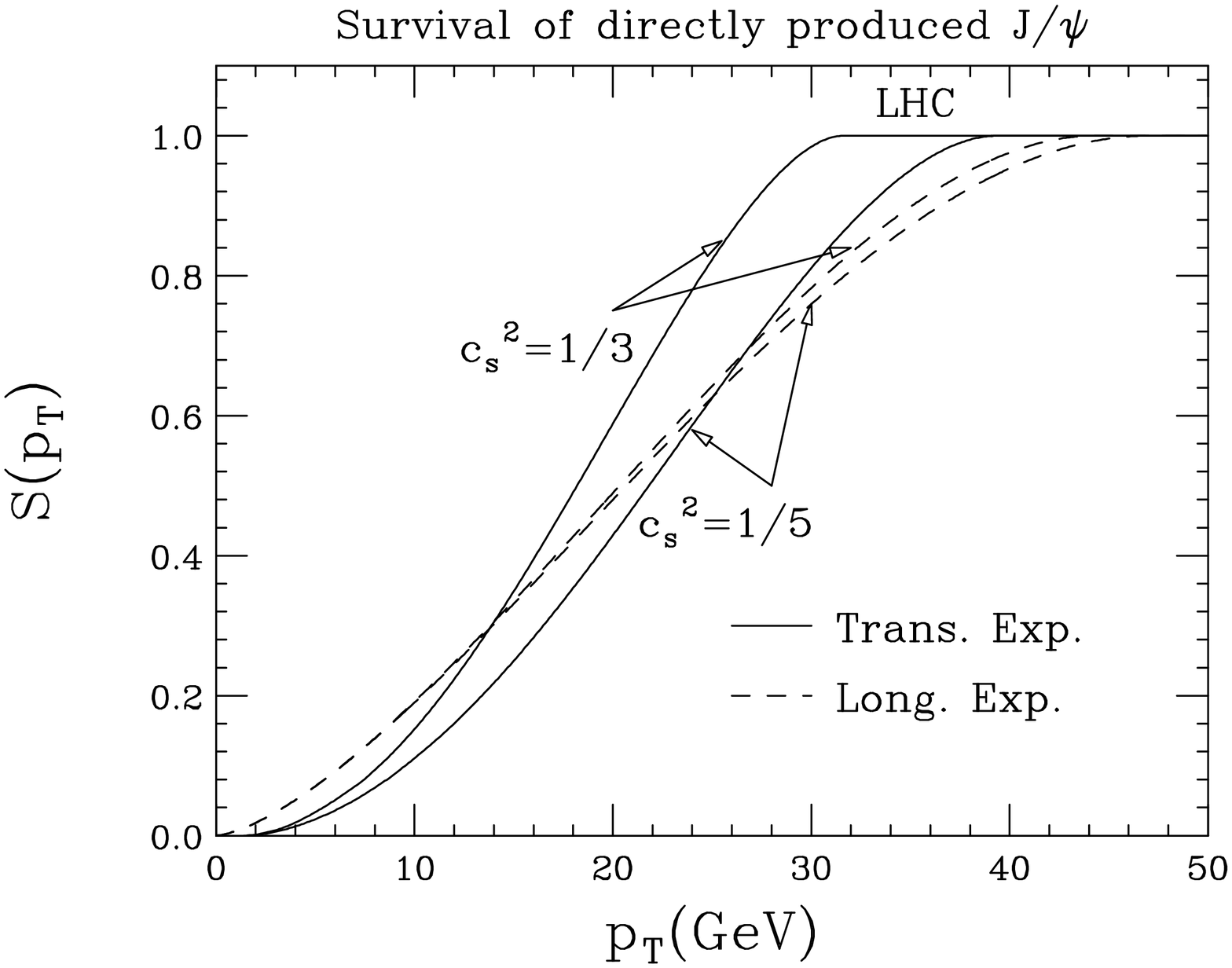,angle=90,height=12cm,width=15cm}
\vskip 0.1in
\caption{The survival probability for $J/\psi$ at  LHC energy
with longitudinal and transverse cooling when speed
of sound is changed.
}
\end{figure}
\newpage
\begin{figure}
\psfig{file=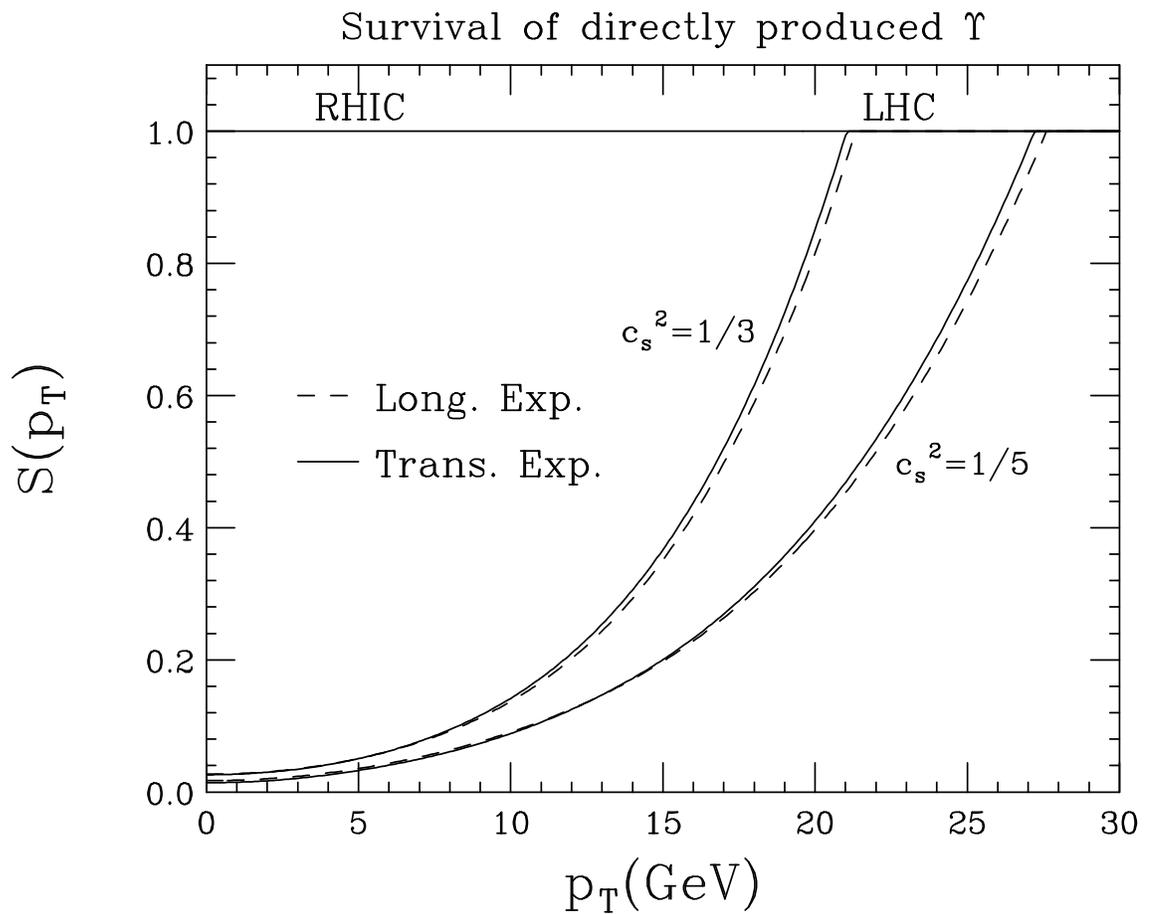,angle=90,height=12cm,width=15cm}
\vskip 0.1in
\caption{The survival probability for $\Upsilon$ at (RHIC and) LHC energy
with longitudinal and transverse cooling when speed
of sound is changed.
}
\end{figure}
\newpage
\begin{figure}
\psfig{file=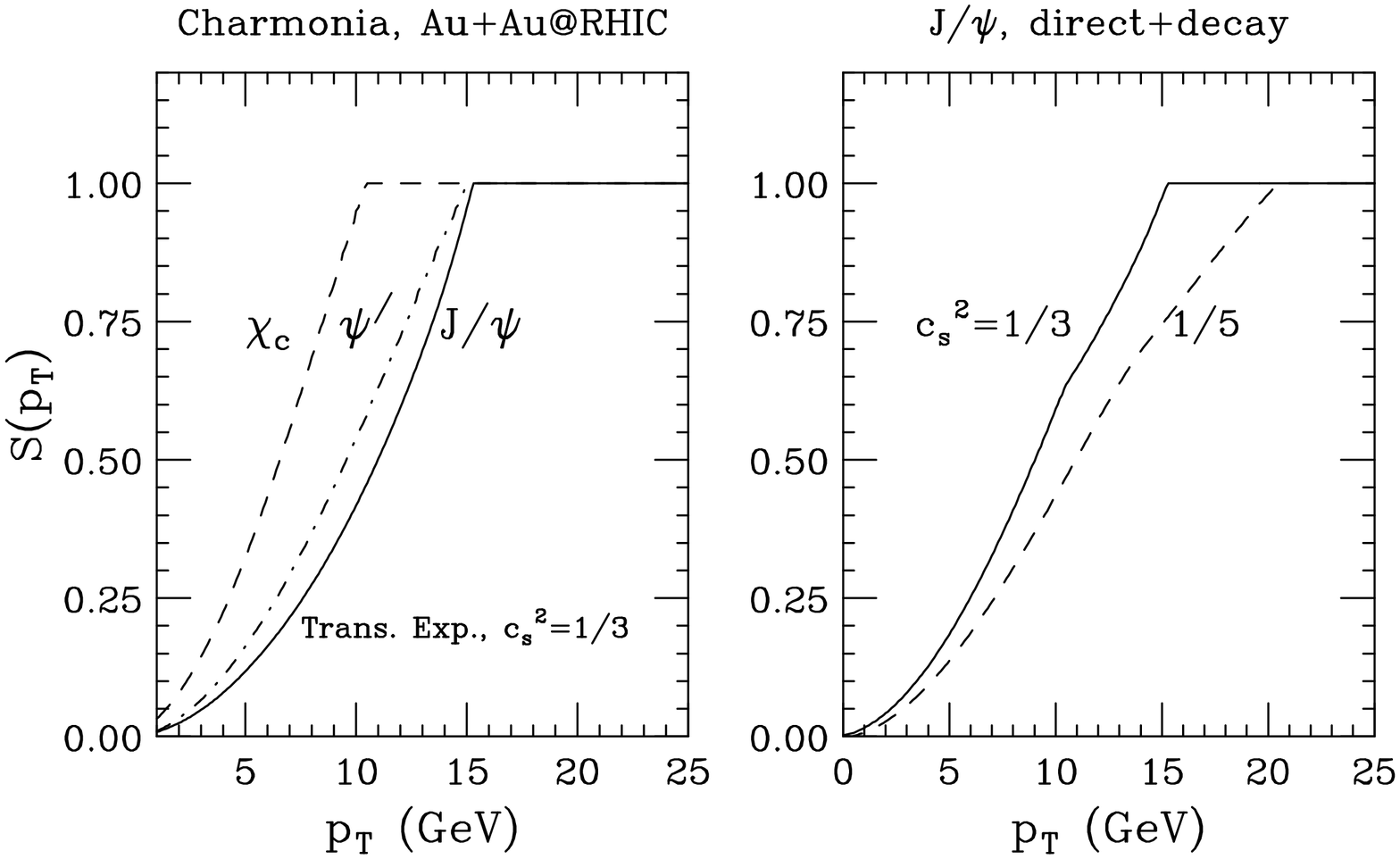,angle=90,height=12cm,width=15cm}
\vskip 0.1in
\caption{The survival probability for  directly produced
$J/\psi$, $\chi_c$ and $\psi^\prime$ at RHIC energies,
for a transversely expanding plasma (left panel), for $c_s^2=1/3$.
The right panel shows the results when the decays of the
resonances is accounted for. 
}
\end{figure}
\newpage
\begin{figure}
\psfig{file=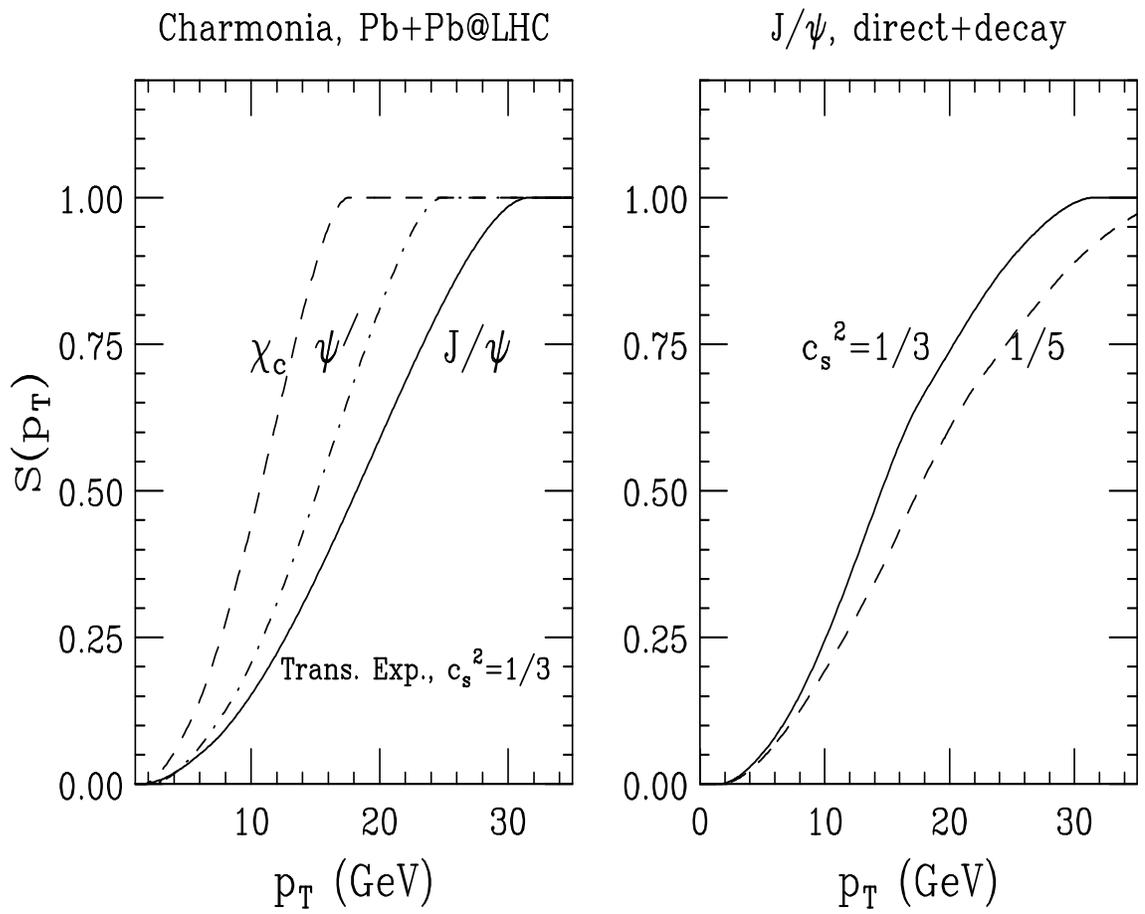,angle=90,height=12cm,width=15cm}
\vskip 0.1in
\caption{Same as Fig.~8, for the survival probability of 
charmonia at LHC energies.
}
\end{figure}
\newpage
\begin{figure}
\psfig{file=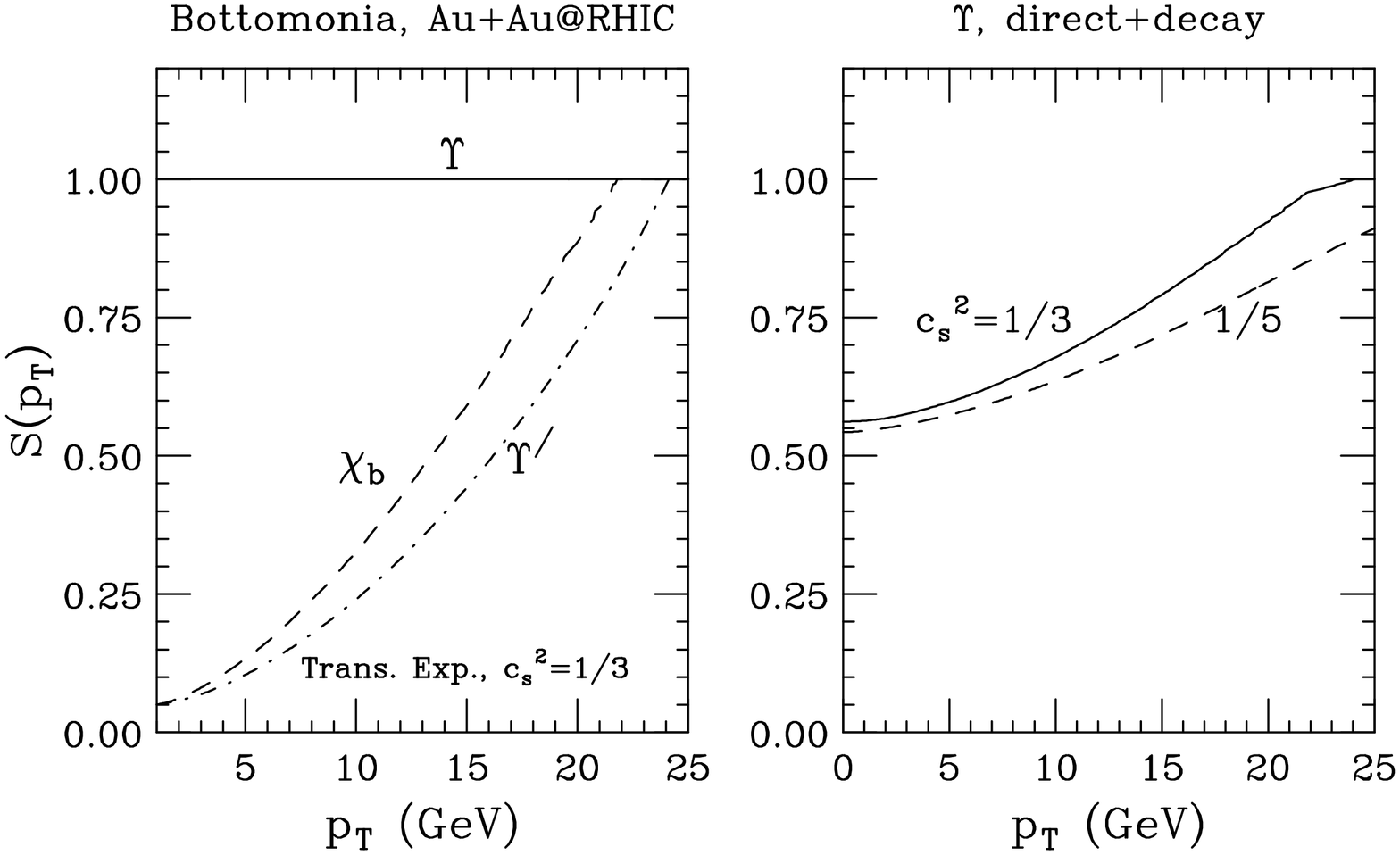,angle=90,height=12cm,width=15cm}
\vskip 0.1in
\caption{The survival probability for  directly produced
$\Upsilon$, $\chi_c$ and $\Upsilon^\prime$ at RHIC energies,
for a transveresly expanding plasma (left panel), for $c_s^2=1/3$.
The right panel shows the results when the decays of the
resonances is accounted for. 
}
\end{figure}
\newpage
\begin{figure}
\psfig{file=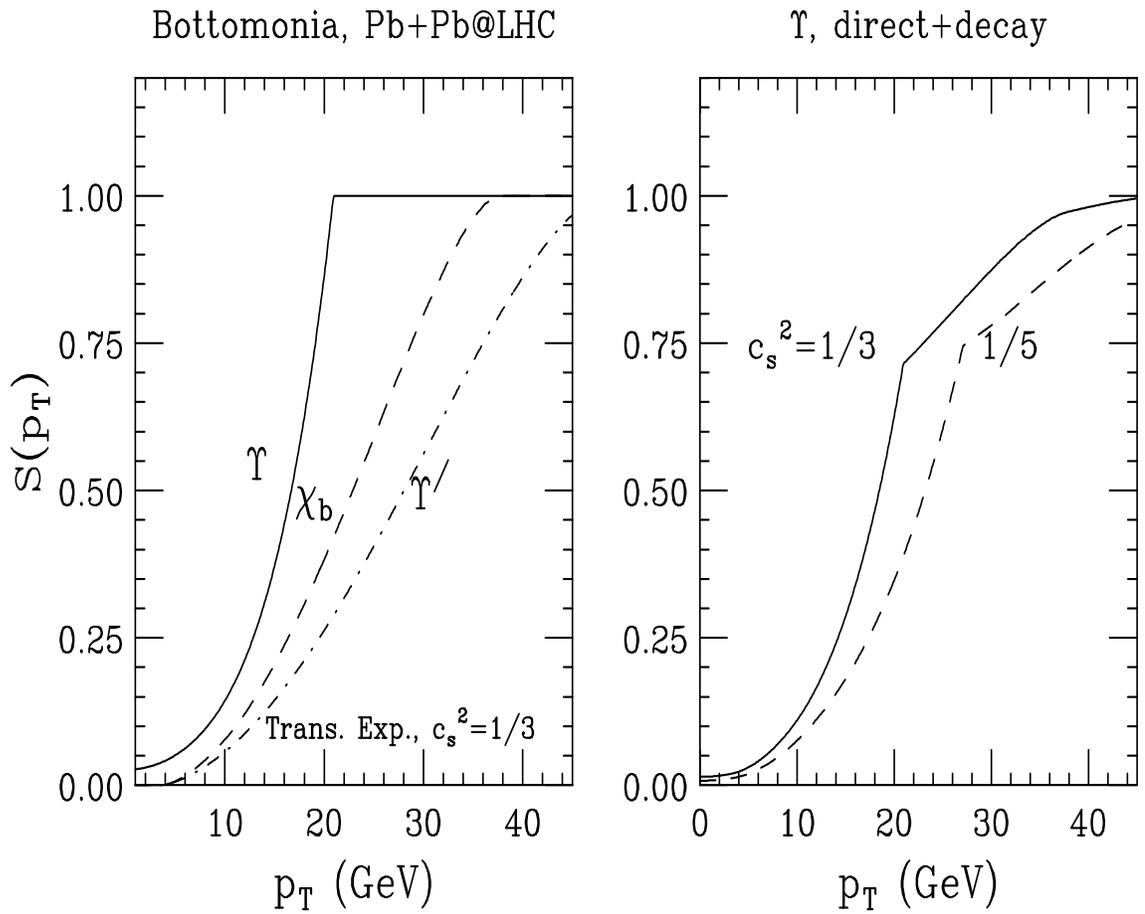,angle=90,height=12cm,width=15cm}
\vskip 0.1in
\caption{Same as Fig.~10, for the survival probability of 
bottomonia at LHC energies.
}
\end{figure}
\end{document}